\documentclass[pdfusetitle,aps,prd,nofootinbib,twocolumn,superscriptaddress,preprintnumbers]{revtex4-1}
\pdfoutput=1
\usepackage{amssymb}
\usepackage{amsmath,mathrsfs}
\usepackage{epsfig}
\usepackage{hyperref}
\usepackage{breakurl}
\usepackage{bm}
\usepackage{color}
\usepackage{tikz}
\usepackage[utf8]{inputenc}
\usepackage[normalem]{ulem}
\usetikzlibrary{decorations.markings}
\usepackage{tikz}
\usetikzlibrary{decorations.pathmorphing,arrows.meta}

\allowdisplaybreaks

\makeatletter
\def\simgt{\mathrel{\lower2.5pt\vbox{\lineskip=0pt\baselineskip=0pt
           \hbox{$>$}\hbox{$\sim$}}}}
\def\simlt{\mathrel{\lower2.5pt\vbox{\lineskip=0pt\baselineskip=0pt
           \hbox{$<$}\hbox{$\sim$}}}}

\def\sectionskip{\vskip .2 cm}

\makeatother

\def\spa#1.#2{\left\langle#1\,#2\right\rangle}
\def\spb#1.#2{\left[#1\,#2\right]}
\def\sand#1.#2.#3{%
\left\langle#1{\vphantom1}\right|{#2}\left|#3\right]}%
\def\sandmp#1.#2.#3{%
\left\langle#1{\vphantom1}\right|{#2}\left|#3\right]}%
\def\sandpm#1.#2.#3{%
\left[#1{\vphantom1}\right|{#2}\left|#3\right\rangle}%
\def\sandmm#1.#2.#3{%
\left\langle#1{\vphantom1}\right|{#2}\left|#3\right\rangle}%
\def\sandpp#1.#2.#3{%
\left[#1{\vphantom1}\right|{#2}\left|#3\right]}%

\renewcommand{\imath}{\mathrm{i}}

\def\nn{\nonumber}
\def\Section#1{\noindent {\it #1}}
\newcommand{\be}{\begin{equation}}
\newcommand{\ee}{\end{equation}}

\newcommand{\Fig}[1]{Fig.~\ref{#1}}
\newcommand{\Eq}[1]{Eq.~\eqref{#1}}
\newcommand{\Eqs}[2]{Eqs.~\eqref{#1} and \eqref{#2}}

\def\rd{ {\rm d}}

\newcommand{\calT}{\mathcal{T}}
\newcommand{\calJ}{\mathcal{J}}
\newcommand{\JiniCM}{\mathsf{J}_{\textrm{CM}}}
\newcommand{\JiniRest}{\mathsf{J}_{\textrm{rest}}}
\newcommand{\Jmatter}{J}

\newcommand{\normd}{\hat{d}}
\newcommand{\normdelta}{\hat{\delta}}

\newcommand{\amp}{\mathcal{M}}
\newcommand{\tensorAA}[1]{#1^{\mu} #1^{\nu}}
\newcommand{\tensorAB}[2]{#1^{(\mu} #2^{\nu)}}

\newcommand{\formF}{\mathcal{F}}
\newcommand{\formG}{\mathcal{G}}
\newcommand{\formH}{\mathcal{H}}

\def\topbotatom#1{\hbox{\hbox to 0pt{$#1\bot$\hss}$#1\top$}} 

\newcommand{\tabeq}[2]{ \parbox{#1}{  \be\begin{aligned}#2 \end{aligned} \nonumber \ee }}

\begin{document}

\title{Radiated Angular Momentum and Dissipative Effects in Classical Scattering}

\author{Aneesh V.~Manohar}
\affiliation{
Department of Physics 0319, University of California at San Diego,
9500 Gilman Drive, La Jolla, CA 92093, USA}
\author{Alexander K.~Ridgway}
\affiliation{
	Department of Physics 0319, University of California at San Diego,
	9500 Gilman Drive, La Jolla, CA 92093, USA}
\author{Chia-Hsien Shen}
\affiliation{
	Department of Physics 0319, University of California at San Diego,
	9500 Gilman Drive, La Jolla, CA 92093, USA}

\begin{abstract}
We present a new formula for the angular momentum $J^{\mu\nu}$ carried away by gravitational radiation in classical scattering. This formula, combined with the known expression for the radiated linear momentum $P^\mu$, completes the set of radiated Poincare charges due to scattering. We parametrize $P^\mu$ and $J^{\mu\nu}$ by non-perturbative form factors and derive exact relations using the Poincare algebra. There is a contribution to $J^{\mu\nu}$ due to static (zero-frequency) modes, which can be derived from Weinberg's soft theorem. Using tools from scattering amplitudes and effective field theory, we calculate the radiated $J^{\mu\nu}$ due to the scattering of two spinless particles to third order in Newton's constant $G$, but to all orders in velocity. Our form-factor analysis elucidates a novel relation found by Bini, Damour, and Geralico between energy and angular momentum loss at $\mathcal{O}(G^3)$. Our new results have several nontrivial implications for binary scattering at $\mathcal{O}(G^4)$. 
We give a procedure to bootstrap an effective radiation reaction force from the loss of Poincare charges due to  scattering.

\end{abstract}
   
\maketitle

\Section{Introduction.}
It is crucial to have accurate theoretical modeling of binary coalescence, given the rapid improvement in sensitivity of current and future gravitational-wave detectors.
Recently,  there has been tremendous progress in solving binary dynamics by utilizing tools in quantum field theory (QFT) building on the pioneering work of nonrelativistic general relativity~\cite{Goldberger:2004jt}.

The power of QFT-based methods originates from the gauge invariance and Lorentz covariance of scattering observables, which can be extracted from QFT amplitudes via effective field theory (EFT) methods~\cite{Goldberger:2004jt,Neill:2013wsa,Cheung:2018wkq}  and the Kosower-Maybee-O'Connell (KMOC) framework~\cite{Kosower:2018adc,Cristofoli:2021vyo}. This enables tools developed in particle physics to be applied to classical gravity. Scattering results can then be translated into binary bound state ones through the effective-one-body mapping~\cite{Buonanno:1998gg,Damour:2016gwp,Damour:2017zjx}, EFT method~\cite{Neill:2013wsa,Cheung:2018wkq} and analytic continuation~\cite{Kalin:2019rwq,Kalin:2019inp,Cho:2021arx}. Outputs from this program naturally fit within the post-Minkowskian (PM) framework, which expands in $G$ but keeps all orders in velocity.
State-of-the-art results for the conservative PM potential~\cite{Bern:2019nnu,Bern:2019crd,Bern:2021dqo,Dlapa:2021npj} and scattering tail effect~\cite{Bern:2021yeh,Dlapa:2021vgp} illustrate the power of this new methodology.

Dissipation is a key feature of binary coalescence that is already present  at 2.5 post-Newtonian (PN) order, as can be seen by the radiation reaction (RR)~\cite{Burke:1970wx,Thorne:1969rba,Chandrasekhar:1970fl,Damour:1981bh,Damour:1982CRAcad,Damour:1983}. The RR force has been extended to up to 4.5PN accuracy ~\cite{Blanchet:1984wm,Jaranowski:1996nv,Nissanke:2004er,Iyer:1993xi,Iyer:1995rn,Blanchet:2018yqa,Galley:2009px,Galley:2012qs,Gopakumar:1997ng}. Theoretical predictions for dissipative effects on binary scattering are also relatively less developed.
For instance, the waveform~\cite{Goldberger:2016iau,Jakobsen:2021smu,Mougiakakos:2021ckm}, impulses~\cite{DiVecchia:2020ymx,DiVecchia:2021bdo,Bjerrum-Bohr:2021din,Damgaard:2021ipf,Brandhuber:2021eyq,DiVecchia:2021ndb,Heissenberg:2021tzo,Herrmann:2021lqe,Herrmann:2021tct} and radiated linear~\cite{Herrmann:2021lqe,Herrmann:2021tct,Riva:2021vnj} and angular momentum~\cite{Damour:2020tta,Bini:2021gat,Gralla:2021qaf,Jakobsen:2021smu,Mougiakakos:2021ckm},
have been computed to only the leading PM order.

The aim of this Letter is to leverage Poincare symmetry to incorporate dissipation due to radiation into the QFT-based framework.  Poincare invariance imposes conservation laws that relate
the linear and angular momentum carried away by radiation, $P^\mu$ and $J^{\mu\nu}$,  to the corresponding loss in the binary system. While the formula for $P^\mu$ is well-known (see also its expression in the KMOC form), the standard formula for $J^{\mu\nu}$~\cite{Maggiore:2007ulw} is less well-understood in scattering scenarios. This is due to the presence of the static mode, which is analogous to the Coulomb mode in electrodynamics (EM)~\cite{Bonga:2018zlx}. In this paper, we derive a new formula~\eqref{eq:PJ_spin2} for $J^{\mu\nu}$ in terms of the stress-energy pseudotensor that applies to radiation with arbitrary frequency. The formula manifests the gauge independence and Lorentz covariance of $J^{\mu\nu}$. This enables us to parametrize $P^{\mu}$ and $J^{\mu\nu}$ with non-perturbative form factors in \Eq{eq:form_factors} that obey exact constraints imposed by the Poincare algebra. Applying this framework perturbatively in $G$, we calculate $J^{\mu\nu}$ to $\mathcal{O}(G^3)$ in \Eqs{eq:J12_G3_full}{eq:J12_G3_full_CM}, and find agreement with the literature~\cite{Damour:2020tta,Bini:2021gat,Gralla:2021qaf}. In particular, we directly derive the remarkable relation~\eqref{eq:BDG_relation} between energy and angular momentum loss first found by~\cite{Bini:2021gat}. Weinberg's soft theorem~\cite{Weinberg:1965nx} greatly simplifies the calculation of the zero-frequency contribution to $J^{\mu\nu}$. Our results, however, disagree with those calculated using standard formula in the rest frame~\cite{Jakobsen:2021smu,Mougiakakos:2021ckm}, due to the subtlety in the static mode.

The radiated Poincare charges have important implications for dissipative binary dynamics. By combining our $\mathcal{O}(G^3)$ results for $J^{\mu\nu}$ with those for $P^\mu$~\cite{Herrmann:2021lqe,Herrmann:2021tct}, one can immediately predict the linear-in-RR correction to the scattering angle and transverse impulse at $\mathcal{O}(G^4)$ using the Bini-Damour formula~\cite{Bini:2012ji,Damour:2020tta} and the maps in~\cite{Bini:2021gat}. In addition, by following the framework in~\cite{Bini:2012ji,Damour:2020tta,Bini:2021gat},  we bootstrap an effective PM RR force via the balance equations~\cite{Iyer:1993xi,Iyer:1995rn,Blanchet:2018yqa} modulo total time derivatives, i.e.~the so-called Schott terms~\cite{Schott:1915zl,Saketh:2021sri}.

\sectionskip
\Section{Radiated Linear and Angular Momentum.}
Consider a scattering process where the initial state consists of massive particles (referred to as matter), and the final state consists of matter and outgoing gravitational radiation.
Poincare symmetry implies that the loss of linear and angular momentum of matter is equal to that carried away by radiation. The radiated linear and angular momentum in the final state are given by
\begin{align}
	P^{\mu} &= \int \rd^3 x \, T^{\mu0}, &
	J^{\mu\nu} &= \int \rd^3 x\,
	x^{[\mu} T^{\nu]0},
	\label{eq:PJ_general}
	\vspace{-10pt}
\end{align}
where $T^{\mu\nu}$ is the stress-energy tensor of the radiation, $a^{[\mu} b^{\nu]} \equiv a^\mu b^\nu - a^\nu b^\mu$, and the integrals are over all space at a fixed time.
The global conserved charges are invariant under improvement terms in $T^{\mu\nu}$~\cite{Callan:1970ze}.

Gravitational radiation is defined as the fluctuation around flat space $g_{\mu\nu} = \eta_{\mu\nu} + \sqrt{32\pi G}\, h_{\mu\nu}$. In what follows, we use the mostly-minus convention. Asymptotically, the radiation can be decomposed into on-shell plane waves labeled by $k^\mu = (\omega,\bm k)$, where $\omega$ is the energy, boldface $\bm k$ denotes spatial momentum, and $k^2 = 0$. 
After gauge fixing, the radiation field of any frequency can be solved in terms of the stress-energy pseudotensor $\calT^{\rho\sigma}(k)$, which is analogous to the current in EM.
$\calT^{\rho\sigma}(k)$ contains both matter and radiation contributions, unlike the usual stress-tensor which does not contain radiation. $\calT^{\rho\sigma}(k)$ is conserved on shell, i.e. $k_\rho \calT^{\rho\sigma}(k)=0$. (One can always find such a $\calT^{\rho\sigma}(k)$~\cite{Kosmopoulos:2020pcd}.)   There is an invariance under residual gauge transformations $\calT^{\mu\nu}(k) \rightarrow \calT^{\mu\nu}(k) + k^\mu \epsilon^\nu(k) + k^\nu \epsilon^\mu(k)$ where $\epsilon(k)\cdot k=0$. 
Crucially, the formula for the radiation field and $J^{\mu\nu}$ are written in terms of $\calT^{\rho\sigma}(k)$, rather than the transverse-traceless components of the radiation field used in the standard formula~\cite{Maggiore:2007ulw}, which means our results are also applicable to the static mode that contributes to the angular momentum. This is similar to the Coulomb field in EM, which is not in the transverse projection of the vector potential.
The radiation field in terms of $\calT^{\rho\sigma}(k)$ reads
\begin{align}
	h_{\mu\nu}(x) &= \sqrt{8\pi G}\int \widetilde{\rd k} \left(
	P_{\mu\nu\rho\sigma}\, \calT^{\rho\sigma}(k)\, e^{-ik\cdot x} +
	\textrm{c.c.}
	\right),
	\label{eq:spin2_mode}
\end{align}
where $P_{\mu\nu\rho\sigma}$ is the gauge-dependent projection and $\widetilde{\rd k}= \frac{ \rd^{3} \bm k } { (2\pi)^{3} 2\omega }$ is the Lorentz invariant phase space measure.

Using Einstein's equations, it is straightforward to relate $T^{\mu\nu}$ to the radiation field. Combining Eqs.~\eqref{eq:PJ_general}, \eqref{eq:spin2_mode}, and  the expression for $T^{\mu\nu}$ in terms of $h_{\mu\nu}(x)$, we obtain the main formulae of this paper,
\begin{align}
	P^{\mu} &= 8\pi G \int \widetilde{\rd k} \, k^\mu
	\left(\calT^{*\rho\sigma}(k) \calT_{\rho\sigma}(k)
	-\frac{1}{2}\calT^{*\rho}_{\rho}(k)\calT^{\sigma}_{\sigma}(k)
	\right), 
	\nn \\
	J^{\mu\nu} &= 8\pi G \int \widetilde{\rd k}
	\Big(
 	\calT^{*\rho\sigma}(k) \mathcal{L}^{\mu\nu}
	\calT_{\rho\sigma}(k)
	-\frac{1}{2}\calT^{*\rho}_{\rho}(k) \mathcal{L}^{\mu\nu}\calT^{\sigma}_{\sigma}(k)
	\nn \\
	&\;\qquad \qquad\qquad +2i \,\calT^{*\rho[\mu}(k) \calT^{\nu]}_{\quad\rho}(k)
	\Big), 	\label{eq:PJ_spin2}
\end{align}
where $\mathcal{L}^{\mu\nu}\equiv i k^{[\mu}\partial^{\nu]}$. Note the absence of any explicit time dependence. This completes the set of expressions for the radiated Poincare charges. Since the stress-energy pseudotensor can be derived directly from on-shell amplitudes using the KMOC framework~\cite{Kosower:2018adc,Cristofoli:2021vyo}, our formulation for $J^{\mu\nu}$ meshes well with the QFT-based approach.
Analogous formul\ae\ for $P^\mu$ and $J^{\mu\nu}$ in EM are given in Appendix~\ref{app:J_EM}.

The expressions in \Eq{eq:PJ_spin2} are highly constrained by gauge invariance and the Poincare algebra. The relative factor between the first and last terms in the $J^{\mu\nu}$ integrand is fixed by invariance under residual gauge transformations.  These terms are sometimes referred to as the orbital and spin contributions. However, only their combination is gauge invariant, implying that individually, they have no physical meaning~\cite{Jaffe:1989jz}. The Poincare algebra imposes the following transformations under the translation $x^\mu \rightarrow x^\mu + a^\mu$,
\begin{align}
	P^\mu &\rightarrow P^{\mu}, \nn \\
	J^{\mu\nu} &\rightarrow J^{\mu\nu} + a^{[\mu} P^{\nu]}.
	\label{eq:translation}
\end{align}
Since $\calT^{\mu\nu}(k) \rightarrow \calT^{\mu\nu}(k) e^{ik\cdot a}$ under translations, the expressions in \Eq{eq:PJ_spin2} indeed obey~\Eq{eq:translation}.

\begin{figure}[t]
	\begin{center}
\includegraphics[]{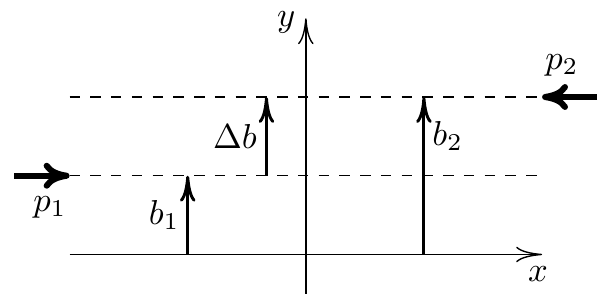}

	\end{center}
	\vskip -.5cm
	\caption{
		The initial configuration of the binary system. The spatial momenta of the two particles are along the $x$ direction, and the impact vectors $b^\mu_{1,2}$ are along the $y$ direction.
	}
	\vspace{-0.4cm}
	\label{fig:coordinates}
\end{figure}

\sectionskip
\Section{Form Factor Parametrization.}
By parametrizing $P^\mu$ and $J^{\mu\nu}$ in terms of the initial data of binary scattering, one can derive additional constraints on them using the Poincare algebra.
The particles are labeled by a Roman subscript $i = 1,2$ and $m_i$, $p^\mu_i$,  and $b^\mu_i$ correspond to the particle's mass, initial momentum and impact vector, which, as depicted in \Fig{fig:coordinates}, obey $p_i \cdot b_j = 0$. In addition, it is useful to define the relative impact vector $\Delta b^\mu \equiv b^\mu_1-b^\mu_2$, its magnitude $b\equiv \sqrt{-\Delta b^2}$, and $\bar{b}^\mu \equiv ( p_1\cdot (p_{1}+p_2)\,b^\mu_1 + p_2\cdot (p_{1}+p_2)\, b^\mu_2)/(p_{1}+p_2)^{2}$. The Lorentz-invariant variables are then $m_i$, $b$, and the relative boost $\sigma \equiv p_1\cdot p_2/(m_1 m_2)$.

We find that the most general forms of $P^\mu$ and $J^{\mu\nu}$ consistent with Lorentz covariance, the Poincare constraints~\eqref{eq:translation}, and particle interchange symmetry are
\begin{align}
	P^\mu &= \formF_1 p^\mu_1 + \formF_2 p^\mu_2 + \formF_3 \Delta b^\mu, \nn \\
	J^{\mu\nu} &= 
	\bar{b}^{[\mu} \left(\formF_1 p_1^{\nu]} + \formF_2  p_2^{\nu]} + \formF_3 \Delta b^{\nu]} \right) \label{eq:form_factors} \\
	&+\Delta b^{[\mu} \left(\formG_1 p_1^{\nu]} - \formG_2 p_2^{\nu]} \right)  
	+\formH_{12} \, p^{[\mu}_2 p^{\nu]}_1, \nn
\end{align}
where $\formF_i,\formG_i,\formH_{12}$ are form factors that are functions of the Lorentz invariants $m_1,m_2,\sigma,b$.  Particle interchange symmetry implies that the form factors satisfy
\begin{align}
	\formF_1 &\stackrel{m_1 \leftrightarrow m_2}{=} \formF_2,  & \formG_1 &\stackrel{m_1 \leftrightarrow m_2}{=} \formG_2, \nn \\
	\formF_3  & \stackrel{m_1 \leftrightarrow m_2}{=}  -\formF_3,  & \formH_{12} &\stackrel{m_1 \leftrightarrow m_2}{=} -\formH_{12},
	\label{eq:conditions_2}
\end{align}
so that the only independent ones are $\formF_2,\formF_3,\formG_2,\formH_{12}$.

We consider two frames in this paper, the center-of-mass (CM) and the frame where particle 1 is initially at rest (referred to as the rest frame hereafter).
See Appendix~\ref{app:frame_choices}  
for the initial conditions in each frame.
In particular, $\bar{b}^\mu=0$ in the CM frame and $b^\mu_1=0$ in the rest frame.
We denote the components of $J^{\mu\nu}$ in the CM and rest frames as $J^{\mu\nu}_{\rm CM}$ and $J^{\mu\nu}_{\rm rest}$, and the initial angular momentum along the $z$ direction as $\JiniCM$ and $\JiniRest$. Their form factor expressions are summarized in~\Eqs{eq:J_CM_general}{eq:J_rest_general}. 
Remarkably, all form factors can be fixed with only $P^\mu$, $J^{12}_{\rm rest}$ and $J^{01}_{\rm rest}$.

Since \Eq{eq:form_factors} was derived from exact symmetries, we can deduce non-perturbative relations among the components of $J^{\mu\nu}_{\rm CM}$ and $J^{\mu\nu}_{\rm rest}$.
For instance, 
this implies $J^{01}_{\rm CM} = J^{01}_{\rm rest}$. In addition, the zero-frequency sector of the radiation has vanishing $\formF_i$ and, according to \Eq{eq:form_factors}, we find
\begin{align}
	\frac{J^{12}_{\rm CM}}{\JiniCM}\bigg|_{\omega= 0} &= \formG_1+\formG_2 \,, \quad
	\frac{J^{12}_{\rm rest}}{\JiniRest}\bigg|_{\omega= 0} = \formG_2 \,.
	\label{eq:ratio_formfactors}
\end{align}
Since the radiation carries no energy at $\mathcal{O}(G^2)$, \Eq{eq:ratio_formfactors} is indeed the full result at this order. As we discuss further below, our formula~\eqref{eq:PJ_spin2} agrees with this relation, but the standard formula~\cite{Maggiore:2007ulw,Jakobsen:2021smu,Mougiakakos:2021ckm} does not.

The remainder of the letter is devoted to using this formalism to compute $J^{\mu\nu}$ in $G$ expansion, defined in Appendix~\ref{app:perturbation_def}.

\begin{figure}[t]
	\begin{center}
\includegraphics[]{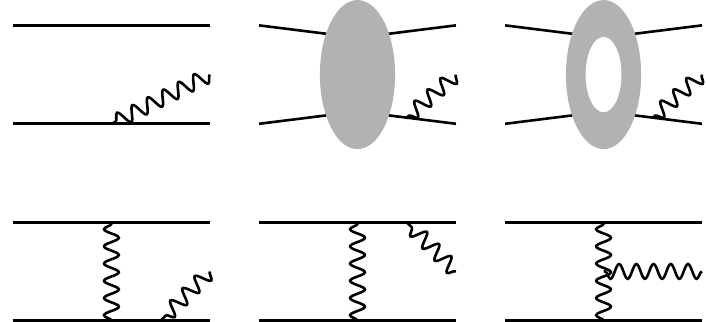}
	\end{center}
	\vskip -.5cm
	\caption{
	Sample diagrams depicting contributions to $\calT^{\mu\nu}(k)$. Straight and wavy lines denote matter and gravitons. Radiation can be emitted from any of the external matter legs. The top row illustrates relevant diagrams for the zero-frequency limit in \Eq{eq:waveform_kink}. The solid and hallow blobs correspond to the $\mathcal{O}(G)$ and $\mathcal{O}(G^2)$ deflections. The bottom row depicts $\calT^{\mu\nu}(k)$ at $\mathcal{O}(G)$ with general frequency.
	}
	\vspace{-0.4cm}
	\label{fig:waveform}
\end{figure}

\sectionskip
\Section{Stress-Energy Pseudotensor.}
To obtain the $\mathcal{O}(G^{3})$ correction to $J^{\mu\nu}$, it is necessary to determine $\calT^{\mu\nu}(k)$ to $\mathcal{O}(G^{2})$.
The diagrams in Fig.~\ref{fig:waveform} depict these contributions. The full expression for $\calT^{\mu\nu}(k)$ is only known up to $\mathcal{O}(G)$~\cite{Goldberger:2016iau,Cristofoli:2021vyo} and is reviewed in Appendix~\ref{app:waveform}.
However, at $\mathcal{O}(G^{2})$ only the related integrand has been constructed~\cite{Shen:2018ebu,Carrasco:2021bmu}.

Fortunately, the full $\mathcal{O}(G^{2})$ expression for $\calT^{\mu\nu}(k)$ is not needed to compute $J^{\mu\nu}$ at $\mathcal{O}(G^3)$, since this term only enters through an interference term with the leading $\mathcal{O}(1)$ static piece. This implies we only need to consider the leading soft limit of $\calT^{\mu\nu}(k)$
which is governed by Weinberg's soft theorem~\cite{Weinberg:1965nx}
\begin{align}
	&\calT^{\mu\nu}(k)|_{\omega\rightarrow 0^+}
	= -\frac{i}{2} \sum_{a=1,2} \frac{p_{a}^\mu p_{a}^\nu}{E_{a}-\hat{\bm k} \cdot \bm p_{a} } 2\pi{\delta}(\omega) 
	\label{eq:waveform_kink} \\
	&+\frac{1}{\omega+i0} \sum_{a=1,2} \,
	\left(
	\frac{p_{a,f}^{\mu} p_{a,f}^{\nu}}{E_{a,f} - \hat{\bm k} \cdot \bm p_{a,f}}
	-\frac{p_{a}^\mu p_{a}^\nu}{E_{a} - \hat{\bm k} \cdot \bm p_{a}}
	\right),\nn
\end{align}
where $\hat{\bm k} = \bm k /\omega$ is a unit vector, and $p^\mu_a =(E_a,\bm p_a)$ and $p^\mu_{a,f} =(E_{a,f},\bm p_{a,f})$ are the initial and final momenta of particle $a$.
The first line is the static contribution to $\calT^{\mu\nu}$ sourced by the initial particles and is exactly soft. The factor of $1/2$ is present to avoid double counting positive- and negative-frequency contributions to the static piece.

The second term in \Eq{eq:waveform_kink} encodes the scattering process.  In the soft limit, the scattering trajectory reduces to a kink at $t=0$ whose frequency space representation is $1/(\omega+i0)$.
The kink is fixed by the hard scattering data, which, for our purposes, is needed to $\mathcal{O}(G^2)$, and can be obtained from the scattering angle $\chi$ summarized in Appendix~\ref{app:angle}.
We also derive \Eq{eq:waveform_kink} in the KMOC framework~\cite{Cristofoli:2021vyo} by taking the soft limit of amplitudes~\cite{Manohar:bs}.
(See also~\cite{Bautista:2019tdr,Bautista:2021llr}.)

\sectionskip
\Section{Perturbative results at $\mathcal{O}(G^2)$.}
At $\mathcal{O}(G^2)$, the only contribution to $J^{\mu\nu}$ is due to interference between the $\mathcal{O}(1)$ static term
and the $\mathcal{O}(G)$ contribution to $\calT^{\mu\nu}$.
As pointed out above and by Damour~\cite{Damour:2020tta}, the zero-frequency limit of $\calT^{\mu\nu}$ is all that is needed for this interference piece, which is given by \Eq{eq:waveform_kink} and the $\mathcal{O}(G)$ impulse.
Evaluating \Eq{eq:PJ_spin2}, and using the notation in Appendix~\ref{app:perturbation_def}
, yields
\begin{align}
	\frac{J^{12}_{\rm rest,2}}{\JiniRest} 
	&=
	\nu M^2 \chi_1\, \mathcal{I}(\sigma)
	=
	\frac{\nu M^2 (2\sigma^2-1)}{\sqrt{\sigma^2-1}}\,
	\mathcal{I}(\sigma) \,, \nn \\
	J^{01}_{\rm rest,2} &=0\,,
	\label{eq:Jrad12_G2}
\end{align}
where $M=m_1+m_2$ is the total mass, $\nu = m_1 m_2 /M^2$ is the reduced mass ratio, 
$\chi_1$ is the $\mathcal{O}(G)$ scattering angle 
given in Appendix~~\ref{app:angle}, and
$\mathcal{I}(\sigma)$ is given in Table~\ref{table:functions}.
Interestingly, $J^{\mu\nu}$ can be written in terms of the scattering angle $\chi_1$ and is independent of  short-distance details.
We find the leading radiated angular momentum is positive when $\chi_1>0$, i.e., the scattering is attractive.

Matching the above to \Eq{eq:form_factors} gives 
\begin{align}
	\formG_{1,2} &= \formG_{2,2}
	= \frac{\nu M^2 (2\sigma^2-1)}{\sqrt{\sigma^2-1}}
	\mathcal{I}(\sigma),
	\label{eq:Gi2}
\end{align}
while all other form factors vanish at this order.  Plugging \Eq{eq:Gi2} into \Eq{eq:form_factors}~ 
gives the remaining components of $J^{\mu\nu}$ in the rest and CM frames:
\begin{align}
	\frac{J^{02}_{\rm CM,2}}{E_1-E_2} &= 
\frac{J^{02}_{\rm rest,2}}{m_1-m_2 \sigma} =
	\frac{b \nu M^2 (2\sigma^2-1)}{\sqrt{\sigma^2-1}}
	\mathcal{I}(\sigma) \,,
	\\
	\frac{J^{12}_{\rm CM,2}}{\JiniCM} 
	&=2  \frac{J^{12}_{\rm rest,2}}{\JiniRest}
	=\frac{2\nu M^2 (2\sigma^2-1)}{\sqrt{\sigma^2-1}}
	\mathcal{I}(\sigma) \,.
	\label{eq:Jrad12_G2_CM}
\end{align}
These results can be directly verified using \Eq{eq:PJ_spin2}.

Our results for $J^{12}_{\rm CM,2}$ agree with Eq.~(4.9) of~\cite{Damour:2020tta}; $J^{02}_{\rm CM,2}$ agrees with Eq.~(160) of~\cite{Gralla:2021qaf} modulo an extra term.
As a nontrivial check, we computed the angular momentum loss using the known 3.5PN RR force exerted on the matter~\cite{Nissanke:2004er}.
This predicts the first two orders of the velocity expansion of $J^{\mu\nu}$ at $\mathcal{O}(G^2)$, and we fully agree in the CM and rest frames.
However, we disagree with the expression for $J^{12}_{\rm rest,2}$ obtained by using the standard formula~\cite{Jakobsen:2021smu,Mougiakakos:2021ckm}, which leads to $J^{12}_{\rm rest,2}/{\JiniRest}= J^{12}_{\rm CM,2}/{\JiniCM}$.
This disagrees with \Eqs{eq:ratio_formfactors}{eq:Jrad12_G2_CM} by a factor of 2 because of the subtlety of applying standard formul\ae\ away from the CM frame~\cite{Bonga:2018gzr}.

The $\mathcal{O}(G^2)$ result is intriguing for a couple reasons. First, it implies that radiation can carry finite angular momentum but, at the same time, zero linear momentum (see also the recent discussions~\cite{Bonga:2018gzr,Ashtekar:2017ydh,Compere:2019gft,Veneziano:2022zwh}).
Second, the ratio $J^{12}/\mathsf{J}$ is different in the CM and rest frames, even in the large mass ratio limit, i.e.~$m_1 \gg m_2$.
Both features can easily be understood using the RR force.
Relative to the binary's initial state, the final state has a smaller impact parameter but the same energy.
This implies the radiated energy vanishes but not the angular momentum.
In addition, the leading 2.5PN RR force exerts the \emph{same} acceleration between the two particles for any mass ratio.
This implies the recoil from the heavy particle cannot be ignored. Therefore, the results do not have to coincide in the CM and rest frames for large mass ratios, in contrast to the conservative effects.

\begin{table}[t]
	\setlength{\tabcolsep}{0pt} 
	\renewcommand{\arraystretch}{3}
	\begin{tabular}{|c|}
		\hline
		\scalebox{0.75}{
		\begin{small}
			\tabeq{10cm}{
				\mathcal{I}(\sigma) &= -\frac{16}{3}+\frac{2\sigma^2}{\sigma^2-1}+\frac{4(2\sigma^2-3)}{\sigma^2-1} \frac{\sigma\, \textrm{arcsinh}\left(\sqrt{\tfrac{\sigma-1}{2}} \right)}{\sqrt{\sigma^2-1}}
				\\
				\frac{\mathcal{E}(\sigma)}{\pi} &= f_1 + f_2 \log\left(\frac{\sigma+1}{2}\right) 
				+f_3 \frac{\sigma\, \textrm{arcsinh}\left(\sqrt{\tfrac{\sigma-1}{2}}\right)}{\sqrt{\sigma^2-1}}
				\\
				\frac{\mathcal{C}(\sigma)}{\pi} &= g_1 + g_2 \log\left(\frac{\sigma+1}{2}\right) 
				+g_3 \frac{\sigma\,\textrm{arcsinh}\left(\sqrt{\tfrac{\sigma-1}{2}} \right)}{\sqrt{\sigma^2-1}}
				\\
				\mathcal{D}(\sigma) &= \frac{3 \pi (5\sigma^2-1)}{8} \mathcal{I}(\sigma)
				\\
				f_1 &= \frac{210\sigma^6 - 552 \sigma^5 + 339 \sigma^4 - 912 \sigma^3 + 
					3148 \sigma^2 - 3336 \sigma + 1151}{48 (\sigma^2-1)^{3/2}}
				\\
				f_2 &= -\frac{35 \sigma^4 + 60 \sigma^3 - 150 \sigma^2 + 76 \sigma - 5}{8 \sqrt{\sigma^2-1}}
				\\
				f_3 &= \frac{(2 \sigma^2 - 3)(35 \sigma^4 - 30 \sigma^2 + 11)}{8 (\sigma^2-1)^{3/2}}
				\\
				g_1 &=\frac{105 \sigma^7- 411 \sigma^6+ 240 \sigma^5 + 537 \sigma^4 - 683 \sigma^3 + 111 \sigma^2 + 386 \sigma -237}{24(\sigma^2-1)^2}
				\\
				g_2 &=\frac{35 \sigma^5- 90 \sigma^4- 70 \sigma^3 + 16 \sigma^2 + 155 \sigma-62}{4(\sigma^2-1)}
				\\
				g_3 &=-\frac{(2\sigma^2-3)(35 \sigma^5 - 60 \sigma^4 - 70 \sigma^3 + 72 \sigma^2 + 19 \sigma-12)}{4(\sigma^2-1)^2}
		}				
		\end{small}	
		}
		\\
		\hline
	\end{tabular}
	\caption{Functions specifying $P^\mu$~\cite{Herrmann:2021lqe,Herrmann:2021tct} and $J^{\mu\nu}$ at $\mathcal{O}(G^2)$ and $\mathcal{O}(G^3)$. These are provided in an ancillary file~\cite{AttachedFile}.
	}
	\label{table:functions}
\end{table}

\sectionskip
\Section{Perturbative results at $\mathcal{O}(G^3)$.}
The $\mathcal{O}(G^3)$ correction to $J^{\mu\nu}$ also contains an interference contribution between the static term from the initial state and the soft limit of $\calT^{\mu\nu}(k)$ in \Eq{eq:waveform_kink}.
At this order, one needs the $\mathcal{O}(G^2)$ correction to the momentum impulses, which can be calculated given the scattering angle up to $\mathcal{O}(G^2)$.

The non-zero frequency contribution to $J^{\mu\nu}$ at $\mathcal{O}(G^3)$ comes from inserting the full $\calT^{\mu\nu}(k)$ at $\mathcal{O}(G)$, reviewed in Appendix~\ref{app:waveform}, into \Eq{eq:PJ_spin2}.  
To evaluate this integral, we worked in the rest frame, expanded the integrand in small velocity, $p_\infty \equiv \sqrt{\sigma^2-1}$, and integrated the resulting terms.
Using this method, we computed the non-relativistic expansion of $J^{12}_{\rm rest,3}$ to $\mathcal{O}(p_\infty^{60})$.
To re-sum the velocity series, we constructed an ansatz by dressing the same transcendental functions appearing in the expression for $P^\mu$ at $\mathcal{O}(G^3)$~\cite{Herrmann:2021lqe,Herrmann:2021tct} with rational functions of $\sigma$ containing unknown coefficients.
The ansatz can be fixed by matching the series to $\mathcal{O}(p_\infty^{54})$ which we confirm to $\mathcal{O}(p_\infty^{60})$.
It would be interesting to verify the resummation using modern integration methods.
As a cross check, we reproduced the results for $P^\mu$ obtained in~\cite{Herrmann:2021lqe,Herrmann:2021tct} using this procedure.

The results for $J^{\mu\nu}$ at $\mathcal{O}(G^3)$ in the rest frame are
\begin{align}
	J^{12}_{\rm rest,3} &= b m_1 m_2^2 \left(
	m_1 \mathcal{C}(\sigma) + (m_1+m_2) \mathcal{D}(\sigma)	\right), \nn \\
	J^{01}_{\rm rest,3} &=
	-b m_1 m_2 (m_1^2-m_2^2)\frac{\chi_1^2}{\sqrt{\sigma^2-1}} \mathcal{I}(\sigma)
	\label{eq:J12_G3_full}
\end{align}
where $\mathcal{C}(\sigma)$ and $\mathcal{D}(\sigma)$ correspond to the non-zero frequency and interference contributions (see Table~\ref{table:functions}).
Using the maps in \Eq{eq:form_factors}
we fix $\formF_{i,3}$, $\formG_{i,3}$ and $\formH_{12,3}$ by matching to $P^{\mu}$~\cite{Herrmann:2021lqe,Herrmann:2021tct} and \Eq{eq:J12_G3_full}:
\begin{align}
	\formF_{2,3} &= 
	\frac{m_1^2 m_2}{\sigma+1} \mathcal{E}(\sigma), \qquad \formF_{3,3} =0\,,
	\nn \\
	\formG_{2,3} &=  \frac{m_1 m_2}{\sqrt{\sigma^2-1}} 
	\Big [
	m_1 \mathcal{C}(\sigma) + (m_1+m_2) \mathcal{D}(\sigma) \nn \\
	& -\frac{(m_2+m_1\sigma)}{M^2 h^2}\,\frac{m_1 m_2 \sqrt{\sigma^2-1} \mathcal{E}(\sigma)}{\sigma+1}
	\Big ] \,,
	\label{eq:form_G3} \\
	\formH_{12,3} &= -b (m_1^2-m_2^2) \frac{\chi_1^2}{\sigma^2-1} \mathcal{I}(\sigma),\nn
\end{align}
where $h = \sqrt{1+2\nu (\sigma-1)}$.

The form factors in \Eq{eq:form_G3} can be used to translate the rest-frame results into CM ones via \Eq{eq:form_factors}.
For instance, 
\begin{align}
	\frac{J^{12}_{\rm CM,3}}{\JiniCM} 
	&=
	\frac{\nu M^3}{p_\infty} 
	\Big [
	\mathcal{C}(\sigma) +2 \mathcal{D}(\sigma) -\frac{\nu p_\infty \mathcal{E}(\sigma)}{h^2}
	\Big ].
	\label{eq:J12_G3_full_CM}
\end{align}
Defining $J_3 \equiv \frac{p_\infty^3}{\nu M^3 h^3}\frac{J^{12}_{\rm CM,3}}{\JiniCM}$,
we find that the combination
\begin{align}
	h^3 J_3 + \frac{\nu p_{\infty}^3}{h^2} \mathcal{E}(\sigma)  = (\sigma^2-1) \left(
	\mathcal{C}(\sigma) +2 \mathcal{D}(\sigma)
	\right)
	\label{eq:BDG_relation}
\end{align}
only depends on $\sigma$ but not $m_{i}$. This is precisely the relation first observed in~\cite{Bini:2021gat}  by considering the matter impulses at $\mathcal{O}(G^4)$.
Here, we obtain the same result as a consequence of the mass scaling at $\mathcal{O}(G^3)$ in \Eq{eq:J12_G3_full} and Lorentz covariance in~\Eq{eq:form_factors}.
Expanding $J_3$ in small velocity yields
\begin{align}
	\frac{J_3}{\pi} &= 
	\frac{28}{5} p_{\infty}^2 +\left(\frac{739}{84}-\frac{163}{15}\nu\right) p_{\infty}^4 \\
	&+\left(-\frac{5777}{2520}-\frac{5339}{420}\nu+\frac{50}{3}\nu^2 \right) p_{\infty}^6 \nn \\
	&+\left(\frac{115769}{126720}+\frac{1469}{504}\nu+\frac{9235}{672}\nu^2 -\frac{553}{24}\nu^3 \right) p_{\infty}^8+\dots. \nn
\end{align}
The first three terms agree with Eq.~(7.21) of~\cite{Bini:2021gat}.

\sectionskip
\Section{Implications for $\mathcal{O}(G^4)$ Scattering.}
It was pointed out in~\cite{Damour:2020tta,Bini:2021gat,Bini:2012ji}  that RR effects on the scattering angle and momentum impulses can be extracted from $P^\mu$ and $J^{\mu\nu}$.
Define the transverse impulse at $\mathcal{O}(G^4)$ to be
$\Delta p_{\perp,4} \equiv \frac{\Delta p_1\cdot \Delta b}{b}\vert_{G^4}$.
It can be written as
\begin{align}
	\Delta p_{\perp,4}	= \nu M^5 \left(G/b\right)^4 (c^{\rm cons}_{b,4}+
	c^{\rm rr,even}_{b,4}
	+ c^{\rm rr,odd}_{b,4}),
\end{align}
where $c^{\rm cons}_{b,4}$ is the conservative contribution calculated in~\cite{Bern:2021yeh,Dlapa:2021vgp}  using the prescription~\cite{Bini:2021gat}, and $c^{\rm rr,even}$ and $c^{\rm rr,odd}$ are the dissipative contributions that are even and odd under time reversal.
($c^{\rm cons}_{b,4}$ and $c^{\rm rr,odd}_{b,4}$ are $c^{\rm cons}_{b,G^4}$ and $c^{\rm rr,tot}_{b,G^4}$ in \cite{Bini:2021gat}.)
Using the explicit map in~\cite{Bini:2021gat}, $c^{\rm rr,odd}_{b,4}$ is fixed by $P^\mu$ and $J^{\mu\nu}$ to $\mathcal{O}(G^3)$
\begin{align}
	c^{\rm rr,odd}_{b,4} &=
	\nu\left[\frac{\sigma(6\sigma^2-5)}{\sigma^2-1}
	-\frac{m_1}{M} \frac{2\sigma^2-1}{(\sigma+1)}\right]\frac{\mathcal{E}(\sigma)}{p_\infty}
	\label{eq:impulse_G4_rr} \\
	&	-\frac{\nu(2\sigma^2-1)}{\sigma^2-1}
	\left[
	\frac{3\pi(5\sigma^2-1)}{2} \mathcal{I}(\sigma)
	+\mathcal{C}(\sigma)+2\mathcal{D}(\sigma)
	\right], \nn
\end{align}
where the mass dependence is consistent with~\cite{Bini:2019nra,Bini:2021gat}.
The first three orders of the velocity expansion in \Eq{eq:impulse_G4_rr} agree with the last line of Eq.~(8.6) of~\cite{Bini:2021gat}.

In the high energy limit $\sigma \rightarrow \infty$, $c^{\rm rr,odd}_{b,4}$ is dominated by terms coming from
$\mathcal{C}(\sigma)$ and $\mathcal{E}(\sigma)$ and scales as $\sigma^3$.
This high-energy behavior is comparable to that of $c^{\rm cons}_{b,4}$.
However, the sum does not cancel and $\Delta p_{\perp,4} \sim G^4 \sigma^3$
in the high energy limit. 
It would be interesting to see if the contribution from $c^{\rm rr,even}_{b,4}$ tames this divergence.

\sectionskip
\Section{Radiation Reaction Force.}
Dissipative effects on a binary system in a generic orbit
can be described by the RR force ${\bm F}_{RR}$.
Let the spinless binary motion lie on the $x-y$ plane.
In polar coordinates, 
$\bm F_{RR} = F_r \bm e_r + \frac{F_\phi}{r} \bm e_\phi$,
where $\bm e_r$ and $\bm e_\phi$ are the radial and angular unit vectors,
$r$ is the relative distance, and $\phi$ is the polar angle.
The energy $E$ and angular momentum $J$ of the binary are not conserved in the presence of ${\bm F}_{RR}$.
Using the formulation in~\cite{Bini:2012ji,Bini:2021gat}, the fluxes of $E$ and $J$ are
\begin{align}
	\dot{\Jmatter} &= F_\phi, \quad
	\dot{E} = \dot{r} F_r + \dot{\phi} F_\phi.
	\label{eq:EOM_RR_EJ}
\end{align}

One can boostrap the RR force using the loss of energy and angular momentum due to scattering in the CM frame.
Assuming ${\bm F}_{RR}$ is a vector under spatial parity and odd under time reversal,
\begin{align}
	{\bm F}_{RR} &=
	c_r p_r\, \bm e_r
	+c_p\, \bm p
	= (c_r+c_p) p_r \bm e_r + \frac{c_p  J}{r} \bm e_\phi
	\label{eq:Frr}
\end{align}
where $\bm p$ is the relative momentum, $p_r= \bm p\cdot \bm e_r$, and $c_r$ and $c_p$ are unknown coefficients that are even under time reversal.
We also assume that $c_r$ and $c_p$ can be expressed in isotropic gauge, i.e. they only depend on $r$ and $\bm p^2$. Classical power counting yields the ansatze
\begin{align}
	c_r = \frac{G^2}{r^3} c_{r,2}\left(\bm p^2 \right) + \dots,\quad 
	c_p = \frac{G^2}{r^3} c_{p,2}\left(\bm p^2 \right) + \dots,
	\label{eq:crr}
\end{align}
where the dots denote higher orders in $G$.

Plugging Eqs.~\eqref{eq:Frr} and \eqref{eq:crr} into \Eq{eq:EOM_RR_EJ},
and integrating over the conservative trajectories in the CM frame, which to leading order are straight lines, yields the change in $E$ and $J$ after scattering
\begin{align}
	\Delta{\Jmatter} &= \frac{2G^2}{b} \frac{E_1 E_2}{E_{12}}\,c_{p,2}\left(\bm p^2_0 \right) + \dots, 
	\label{eq:EJloss} \\
	\Delta{E} &= \frac{2G^2 \JiniCM}{3 b^3} (c_{r,2}\left(\bm p^2_0 \right) + 3 c_{p,2}\left(\bm p^2_0 \right)) + \dots\,, \nn
\end{align}
Conservation of energy and angular momentum implies, 
\begin{align}
	\Delta{\Jmatter} &= -J^{12}_{\rm CM}, \quad
	\Delta{E} = -P^{0}_{\rm CM} \,.
	\label{eq:balance_total}
\end{align}
Matching this to \Eq{eq:Jrad12_G2_CM} and $P^{0}_{\rm CM,2}=0$ at $\mathcal{O}(G^2)$ fixes the ansatze entirely:
\begin{align}
	c_{r,2}\left(\bm p^2_0 \right) &= -3 c_{p,2}\left(\bm p^2_0 \right) ,
	\label{eq:RRforce_G2} \\
	c_{p,2}\left(\bm p^2_0 \right) &=  - \frac{\nu^2 M^4}{E_1 E_2} (2\sigma^2-1)\, \mathcal{I}(\sigma).
	\nn
\end{align}
This extends the RR force at $\mathcal{O}(G^2)$ to all orders in velocity, which was only derived previously to the first three orders in the velocity expansion~\cite{Gopakumar:1997ng}.
The equations of motion for the CM recoil,
and the extension to $\mathcal{O}(G^3)$,
can be studied similarly.

This analysis, however, assumes that ${\bm F}_{RR}$ can be expressed in the isotropic gauge.
It would be important to check the agreement between \Eq{eq:RRforce_G2} with the known RR force in the overlapping region by including Schott terms~\cite{Schott:1915zl}, i.e. total time derivatives that leave $\Delta J$ and $\Delta E$ invariant~\cite{Saketh:2021sri}.
We leave this to future work.

\sectionskip
\Section{Conclusions.} 
In this Letter, we build a new framework to calculate the radiated angular momentum due to scattering that meshes well with QFT-based methods.
Our work opens up many avenues for future work. Some obvious generalizations include dissipative effects in scattering with  spin~\cite{Jakobsen:2021lvp,Jakobsen:2022fcj}, and in gauge~\cite{Saketh:2021sri,Bern:2021xze} and supersymmetric theories~\cite{Amati:1990xe,DiVecchia:2020ymx}.
It would also be interesting to compare our method with other approaches using soft theorems~\cite{DiVecchia:2021ndb,Heissenberg:2021tzo}.
A crucial next step is to calculate $P^{\mu}$ and $J^{\mu\nu}$ to $\mathcal{O}(G^4)$.
For the bounded binaries,
it would be interesting to compare with the flux from analytic continuation~\cite{Cho:2021arx}, and study its impact on waveform models~\cite{Antonelli:2019ytb}.
Last but not least, it would be interesting to extend our framework beyond gravitational-wave science, perhaps along the lines of jet observables~\cite{Basham:1978bw}.

\sectionskip
\Section{Acknowledgments.}                                                                 
We thank Z.~Bern, T.~Damour, W.~Goldberger, D.~O'Connell, J. Plefka, R.~Porto, R.~Roiban, I.~Rothstein, and M.~Solon
for comments on the manuscript, and L.~Blanchet, T.~Damour, E.~Herrmann, D.~Kosower, Z.~Moss, J.~Parra-Martinez, M.~Ruf, and M.~Solon for helpful discussions.
This work is supported in part by the U.S.\ Department of Energy (DOE) under award
number~DE-SC0009919. 
C.-H.S. is also grateful to
Mani L. Bhaumik Institute for Theoretical Physics for hospitality during the completion of this work.

\vskip .3 cm 

\appendix
\section{Radiated Linear and Angular Momenta in Electromagnetism}
\label{app:J_EM}
In this section, we derive the analog of Eq.~(3) of the main paper
~in electromagnetism (EM).
The EM stress-energy tensor associated with the gauge field $A_\mu$ can be written as
\begin{align}
	T^{\mu\nu} = -F^{\mu}\,_{\rho} F^{\nu\rho} + \frac{\eta^{\mu\nu}}{4} F^{\rho\sigma} F_{\rho\sigma},
	\label{eq:T_EM}
\end{align}
where $F_{\mu\nu}=\partial_{[\mu}A_{\nu]}$ is the field strength.
Once a gauge is chosen, $A^\mu$ can be expressed in terms of the source current using the classical Green's function.
In momentum space, the gauge field is given by
\begin{align}
	A_\mu(x) &= \int \widetilde{dk} \left(
	P_{\mu\nu}\, \calJ^\nu(k)\, e^{-ik\cdot x} +
	\textrm{c.c.}
	\right),
	\label{eq:spin1_mode}
\end{align}
where $\calJ^\nu(k)$ is the conserved current satisfying $k_\mu \calJ^\mu(k)=0$,
and $P_{\mu\nu}$ is the gauge-dependent transverse projector.
In Lorentz gauge, one replaces $P_{\mu\nu} \calJ^\nu(k)$ with $\calJ_\mu(k)$ in Eq.~\eqref{eq:spin1_mode}.
Like the stress-energy pseudotensor in gravity, the EM source current exhibits a residual gauge freedom: $\calJ^\mu(k) \rightarrow \calJ^\mu(k) + \alpha k^\mu$.

Combining Eq.~(1)
~of the main text and \Eqs{eq:T_EM}{eq:spin1_mode} above, the EM expressions for $P^{\mu}$ and $J^{\mu\nu}$ are
\begin{align}
	P^{\mu} &= \int \widetilde{dk} \, k^\mu  \left(-\calJ^{*\rho}(k) \calJ_{\rho}(k)\right), 
	\label{eq:PJ_spin1}
	\\
	J^{\mu\nu} &= \int \widetilde{dk} \,
	\big(
	-\calJ^{*\rho}(k)\, \mathcal{L}^{\mu\nu} \calJ_{\rho}(k)
	-i \calJ^{*[\mu}(k) \calJ^{\nu]}(k)
	\big).\nn
\end{align}
Like in gravity, the explicit time dependence and gauge choice drop out and we are left with expressions that only depend on $\calJ^\mu(k)$.
This radiated angular momentum formula is valid for radiation with arbitrary frequency.
As a consistency check, the formul\ae\ \Eq{eq:PJ_spin1} are also invariant under the residual gauge transformation and obey the Poincare algebra.

\section{Frame Choices and Form Factor Expressions}
\label{app:frame_choices}
In this section, we summarize the initial conditions of the matter in the CM and rest frames.  
The two frames are related by a boost along the $x$ axis and a translation along the $y$ axis.
In both frames, $\Delta b^\mu=(0,0,-b,0)$ as depicted in~Fig.~1 of the main text.

In the CM frame, the initial conditions are
\begin{align}
	p^\mu_1 &=(E_1,|\bm p_0|,0,0),& \quad p^\mu_2 &=(E_2,-|\bm p_0|,0,0), \\
	b^\mu_1 &=\frac{E_2}{E_1+E_2} \Delta b^\mu,& \quad b^\mu_2 &=-\frac{E_1}{E_1+E_2}\Delta b^\mu, \nn
\end{align}
such that $\bar{b}^\mu=0$.
Here, $E_i = \sqrt{\bm p_0^2 + m_i^2}$ denotes the initial energy of the scalars.
The initial angular momentum lies along the $z$ direction with magnitude $\JiniCM = |\bm p_0|b$.

In the rest frame, particle 1 is initially at rest and sits at the origin, 
\begin{align}
	p^\mu_1 &=(m_1,0,0,0),& \quad &p^\mu_2=(\sigma m_2,-p_{\infty}m_2,0,0), \\
	b^\mu_1 &=(0,0,0,0),& \quad   &b^\mu_2=-\Delta b^\mu, \nn
\end{align}
such that in this frame
\begin{align}
	\bar{b}^\mu=
	-\frac{m_2(m_2+m_1\sigma)}{M^2 h^2} \Delta{b}^\mu.
\end{align}
Recall that
$h \equiv \left((p_1+p_2)^2/M^2\right)^{\frac{1}{2}} = \sqrt{1+2\nu (\sigma-1)}$.
The magnitude of the initial angular momentum is $\JiniRest = p_{\infty} m_2 b$.

Given the initial conditions, the components of $J^{\mu\nu}$ can be expressed in terms of form factors.
The nontrivial components in the CM frame are
\begin{align}
	J^{12}_{\rm CM} &= \JiniCM (\formG_1 +\formG_2) \,, \nn \\
	J^{02}_{\rm CM} &= b\, (\formG_1 E_1 - \formG_2 E_2)  \,,
	\label{eq:J_CM_general} \\
	J^{01}_{\rm CM} &= m_1 m_2 p_\infty \formH_{12}\,,\nn 
\end{align}
and in the rest frame
\begin{align}
	J^{12}_{\rm rest} &=
	\JiniRest\, \left(\formG_2 +\frac{m_2 (m_2+m_1\sigma)}{M^2 h^2}  \formF_2\right) \,,
	\nn \\
	J^{02}_{\rm rest} &= b\, 
	(\formG_1 m_1 - \formG_2 m_2 \sigma) 
	\label{eq:J_rest_general} \\
	&- \frac{m_2(m_2+m_1\sigma)b}{M^2 h^2} (\formF_1 m_1 + \formF_2 m_2 \sigma) \,, \nn \\
	J^{01}_{\rm rest} &= m_1 m_2 p_\infty \formH_{12}\,.\nn
\end{align}
In particular, we can obtain all form factors by calculating $P^\mu$, $J^{12}_{\rm rest}$ and $J^{01}_{\rm rest}$ and using
the particle exchange symmetry described in the main text.
Evidently, $J^{02}_{\rm rest}$ and the CM frame results become consistency checks.

\section{Perturbative Expansion in $G$}
\label{app:perturbation_def}
The PM expansion organizes terms in powers of $G/b$.  The coefficients are accurate to all orders in the velocity.  We can expand the linear and angular momentum as
\begin{align}
	P^{\mu}_{\rm X} &=  \left(\frac{G}{b}\right)^3 P^{\mu}_{\rm X,3}+\dots\,, \\
	J^{\mu\nu}_{\rm X} &=  \left(\frac{G}{b}\right)^2 J^{\mu\nu}_{\rm X,2}+\left(\frac{G}{b}\right)^3 J^{\mu\nu}_{\rm X,3}+\dots\,,
\end{align}
where $\textrm{X}=\text{CM}$ or $\text{rest}$.
The form factors can also be expanded as
\begin{align}
	\formG_i &= \left(\frac{G}{b}\right)^2 \formG_{i,2}+\left(\frac{G}{b}\right)^3 \formG_{i,3}+\dots \,, 
	\label{eq:form_factors_G} \\
	\formF_i &= \left(\frac{G}{b}\right)^3 \formF_{i,3}+\dots \,, \nn \\
	\formH_{12} &= \left(\frac{G}{b}\right)^3 \formH_{12,3}+\dots \,.\nn
\end{align}
The dots denote higher order corrections in the $G/b$ expansion.

\section{Conservative Scattering Angle}
\label{app:angle}
The conservative scattering angle in the CM frame is
\begin{align}
	\chi &=
	\sum_{a} \left(\frac{G m_1 m_2}{\JiniCM} \right)^a 2\chi_a\,,
	\label{eq:angle_cons} \\
	\chi_1 &=
	\frac{(2\sigma^2-1)}{\sqrt{\sigma^2-1}},\quad
	\chi_2 = \frac{3\pi}{8} \frac{5\sigma^2-1}{\sqrt{1+2\nu(\sigma-1)}} \,.
	\label{eq:angle_cons_G}
\end{align}
The impulses in the CM frame are 
\begin{align}
	p^\mu_{1,f}- p^\mu_1 = -(p^\mu_{2,f}- p^\mu_2)
	&= |\bm p_0| (0,\cos \chi, \sin \chi,0).
\end{align}

\section{Stress-energy Pseudotensor at $\mathcal{O}(G)$}
\label{app:waveform}
The full stress-energy pseudotensor  at $\mathcal{O}(G)$ sourced by binary scattering has been calculated using various methods,
including the worldline and the KMOC framework.
It can be written as
\begin{align}
	\mathcal{T}^{\mu\nu}(k)
	&= i\int \normd \ell\
	\normdelta(2 p_1\cdot l) \normdelta(2 p_2\cdot \ell) e^{i l \cdot b_1 -i\ell \cdot b_2}
	\amp^{\mu\nu}_5(\ell,k) \,,
	\label{eq:connected_waveform}
\end{align}
where $\normd \ell=\frac{d^D\ell}{(2\pi)^D}$ and $l=k+\ell$.
The kernel $\amp^{\mu\nu}_5(\ell,k)$ is closely related to the tree-level amplitude $\amp_5(\ell,k)$ of 
five-particle scattering depicted by the diagrams in the bottom row of Fig.~2 of the main text.

We use the KMOC formalism to extract $\amp^{\mu\nu}_5(\ell,k)$.
First, we strip off the graviton polarization $e_{\mu\nu}$ from $\amp_5(\ell,k)$ 
such that $\amp^{\mu\nu}_5(\ell,k)$ is a symmetric tensor and
reproduces the correct amplitude,
$e_{\mu\nu}\,\amp^{\mu\nu}_5(\ell,k)=\amp_5(\ell,k)$.
Next, we impose that the Ward identity is satisfied even when one index is free,
i.e.~$k_\mu \amp^{\mu\nu}_5(\ell,k) =0$, to
ensure the conservation of the stress-energy pseudotensor, $k_\mu \calT^{\mu\nu}(k) =0$.
This is made possible by using the pure gauge freedom in $\amp^{\mu\nu}_5(\ell,k)$.
Finally,
we isolate the classical contribution to $\amp^{\mu\nu}_5(\ell,k)$ by rescaling $(\ell,k)\rightarrow (\lambda\ell,\lambda k)$ and expanding in small $\lambda$. 
For this purpose, it is sufficient to truncate to $\mathcal{O}(1/\lambda^2)$.
In contrast to the conservative case,
there is actually an iteration contribution if one keeps the Feynman $i\varepsilon$ in the matter propagators.
As expected, this iteration is precisely canceled by the cut contribution in the KMOC approach.
Therefore, we can effectively ignore the Feynman $i\varepsilon$ when taking the classical limit.
This procedure yields
\begin{align}
	\amp^{\mu\nu}_5(\ell, k)
	&=
	8\pi G \left(D^{\mu\nu}_1 + D^{\mu\nu}_2 + D^{\mu\nu}_{3} \right),
	\label{eq:M5}
\end{align}
where
\begin{align}
	D^{\mu\nu}_1 &=
	\frac{2m_1^2 m_2^2 (2\sigma^2-1)}{(p_1\cdot k)^2 \ell^2}
	\left(
	2(p_1 \cdot k) \tensorAB{p_1}{\ell}
	-(\ell \cdot k) \tensorAA{p_1}
	\right) \nn \\
	& +\frac{8m_1 m_2 \sigma}{(p_1\cdot k) \ell^2} 
	\left(
	2(p_1 \cdot k) \tensorAB{p_1}{p_2}
	-(p_2 \cdot k) \tensorAA{p_1}
	\right) 
	\,, \\
	D^{\mu\nu}_2 &=
	\frac{2m_1^2 m_2^2 (2\sigma^2-1)}{(p_2\cdot k)^2 l^{2}}
	\left(
	(\ell \cdot k) \tensorAA{p_2}
	-2(p_2 \cdot k) \tensorAB{p_2}{\ell}
	\right) \nn \\
	& +\frac{8m_1 m_2 \sigma}{(p_2\cdot k) l^{2}} 
	\left(
	2(p_2\cdot k) \tensorAB{p_1}{p_2}
	-(p_1\cdot k) \tensorAA{p_2}
	\right) 
	\,, \\
	D^{\mu\nu}_3 &=
	\frac{16m_1 m_2 \sigma}{\ell^2 l^{2}}
	\left(
	(p_2\cdot k) \tensorAB{p_1}{\ell}
	-(p_1\cdot k) \tensorAB{p_2}{\ell}
	\right) \nn \\
	& -\frac{8}{\ell^2 l^{2}}
	\left(
	(p_1 \cdot k)^2 \, \tensorAA{p_2}
	+(p_2 \cdot k)^2 \, \tensorAA{p_1}
	\right)  \nn \\
	& +\frac{8\tensorAB{p_1}{p_2}}{\ell^2 l^{2}}
	\left(2(p_1 \cdot k)(p_2\cdot k)
	-m_1 m_2 \sigma\left(\ell^2 + l^{2}\right)
	\right) \nn \\
	&-\frac{4m_1^2 m_2^2 (2\sigma^2-1)}{\ell^2 l^{2}}
	\tensorAA{\ell}\,.
\end{align}
We use $a^{(\mu}b^{\nu)}\equiv (a^\mu b^\nu+a^\nu b^\mu)/2$ in the above.
Together with Eq.~(8) of the main text
, this completes the set of expressions for the stress-energy pseudotensor that are needed to compute $J^{\mu\nu}$ to $\mathcal{O}(G^3)$.
\vskip -.3 cm 

\bibliography{jrad}

\end{document}